\documentclass[prl,aps,twocolumn,superscriptaddress,floatfix]{revtex4-2}
\usepackage[latin9]{inputenc}
\usepackage{mathtools}
\usepackage{amsbsy}
\usepackage[caption=false]{subfig}
\usepackage{dsfont}
\usepackage{amstext}
\usepackage[unicode=true,pdfusetitle,
 bookmarks=false,
 breaklinks=false,pdfborder={0 0 1},backref=false,colorlinks=false]
 {hyperref}
\hypersetup{
 bookmarksnumbered=false,bookmarksopen=false}

\usepackage{hyperref}
\hypersetup{colorlinks,citecolor=NavyBlue,filecolor=black,linkcolor=black,urlcolor=black}
\usepackage[dvipsnames]{xcolor}
\usepackage{graphicx}

\makeatletter

\@ifundefined{textcolor}{}
{%
 \definecolor{BLACK}{gray}{0}
 \definecolor{WHITE}{gray}{1}
 \definecolor{RED}{rgb}{1,0,0}
 \definecolor{GREEN}{rgb}{0,1,0}
 \definecolor{BLUE}{rgb}{0,0,1}
 \definecolor{CYAN}{cmyk}{1,0,0,0}
 \definecolor{MAGENTA}{cmyk}{0,1,0,0}
 \definecolor{YELLOW}{cmyk}{0,0,1,0}
}

\newcommand\ket[1]{\left|#1\right\rangle}
\newcommand\bra[1]{\left\langle #1 \right|}
\newcommand{\adag}{a^{\dagger}}
\newcommand{\adaga}{a^{\dagger}a}

\def\<#1>{\mathinner{\langle#1\rangle}}
\def\|#1>{\mathinner{|#1\rangle}}

\usepackage{amsmath}
\usepackage{graphicx}
\usepackage{amssymb}
\usepackage{txfonts,color}
\makeatother

\begin{document}
\title{Multi-squeezed state generation and universal bosonic control via a driven quantum Rabi model}
\author{Peter McConnell} 
\affiliation{Centre for Theoretical Atomic, Molecular and Optical Physics, Queen's University Belfast, Belfast BT7 1NN, United Kingdom}
\author{Alessandro Ferraro}
\affiliation{Centre for Theoretical Atomic, Molecular and Optical Physics, Queen's University Belfast, Belfast BT7 1NN, United Kingdom}
\affiliation{Dipartimento di Fisica Aldo Pontremoli, Universit\`a degli Studi di Milano, I-20133 Milano, Italy}
\author{Ricardo Puebla}
\affiliation{Instituto de F{\'i}sica Fundamental (IFF), CSIC, Calle Serrano 113b, 28006 Madrid, Spain}
\affiliation{Departamento de F{\'i}sica, Universidad Carlos III de Madrid, Avda. de la Universidad 30, 28911 Legan{\'e}s, Spain}

\begin{abstract}
Universal control over a bosonic degree of freedom is key in the quest for quantum-based technologies. Such universal control requires however the ability to perform demanding non-Gaussian gates --- namely, higher-than-quadratic interactions at the level of the bosonic operators. Here we consider a single ancillary two-level system, interacting with the bosonic mode of interest via a  driven quantum Rabi model, and show that it is sufficient to induce the deterministic realization of a large class of Gaussian and non-Gaussian gates, which in turn provide universal bosonic control. This scheme reduces the overhead of previous ancilla-based methods where long gate-sequences are required to generate highly populated targets. In fact, our method naturally yields the high-fidelity preparation of multi-squeezed states --- \textit{i.e.}, the high-order generalization of displaced and squeezed states --- which feature large phase-space Wigner negativities. The universal control is further illustrated by generating a cubic-phase gate. Finally, we address the resilience of the method in the presence of realistic noise. Due to the ubiquity of the considered interaction, our scheme might open new avenues in the design, preparation, and control of bosonic states in different setups.
\end{abstract}

\maketitle

{\em Introduction.---}
The control of quantum systems composed of distinguishable bosons --- also referred to as \textit{quantum continuous variables} --- is currently the subject of an intense research effort in the context of quantum information science and technology \cite{Braunstein:05, Weedbrook:12, serafini2017quantum}. In fact, the infinite-dimensional Hilbert space of bosonic systems can be used to embed logical qubits with a twofold potential benefit: errors can be corrected in a hardware efficient manner \cite{terhal_towards_2020, grimsmo_quantum_2021, fukui_building_2022-1} and highly scalable platforms can be harnessed \cite{bourassa2021blueprint, asavanant_generation_2019, larsen_deterministic_2019, jolin2021multipartite}. Therefore, bosonic systems have been recognized as major contenders in the quest for quantum advantage \cite{zhong2021phase, madsen2022quantum} and fault-tolerant quantum computation \cite{ofek2016extending}. At the core of this capability there are quantum states and dynamics beyond what is known to be simulatable efficiently via classical means \cite{bartlett2002, mari2012, veitch2013, rahimi-keshari2016, garcia-alvarez2020, calcluth2022}. In particular, states that display phase-space negativities, in terms of their associated Wigner functions, have been recognized as a rigorously quantifiable resource \cite{Albarelli:18, takagi2018}. 

The generation of such states and, more in general, arbitrary bosonic control can be achieved via tunable linear and quadratic Hamiltonians, provided an additional higher-order Hamiltonian (of any form) is also at disposal \cite{Lloyd:99}. The latter induces in fact the crucial dynamics that unlocks the generation of Wigner negativities. In practice, the experimental implementation of such operations is challenging --- being hindered by the weakness of natural higher-order terms in bosonic systems --- and various approaches have been pursued to overcome this issue. In the context of optical platforms conditional measurements are typically employed \cite{lvovsky_production_2020}, whereas dissipation engineering has been considered for example in superconducting circuits \cite{Mirrahimi2014, Leghtas15} and optomechanics \cite{rips2012steady, Tan2013a, asjad2014reservoir, brunelli2018unconditional, houhou2022unconditional}. One of the most valid approaches --- which removes the need of a probabilistic strategy or a controlled reservoir --- consists of using an ancillary finite-dimensional system, which effectively induces a unitary dynamics corresponding to higher-order Hamiltonians at the level of the bosonic system alone \cite{ma_quantum_2021}. A variety of platforms can host such a system, including cavity \cite{haroche2006exploring} and circuit QED \cite{blais2021circuit}, trapped ions \cite{Leibfried:03}, nanophotonics \cite{tiecke2014nanophotonic} and optomechanics~\cite{Aspelmeyer:14}. 

Historically, the latter approach was put forward in a seminal contribution by Law and Eberly \cite{law1996arbitrary}, where the Jaynes-Cummings model is used to swap excitations sequentially between an ancillary qubit and the bosonic system of interest. Notwithstanding its successful implementation \cite{hofheinz2009synthesizing}, this model has significant drawbacks in terms of the length of the sequence needed to generate highly populated targets and, more in general, even larger sequences for arbitrary unitary control are required \cite{mischuck2013qudit}. Alternatives were therefore sought --- based on models working in the dispersive rather than in the resonant regime --- where both quantum control \cite{heeres2017implementing} and gate-based strategies \cite{heeres2015cavity, krastanov2015universal} have been introduced and successfully implemented. The former provides faster operations but lacks the modularity and transparency of the latter, and both still have limited applicability when highly populated targets are addressed in noisy settings \cite{kudra2022robust}, due to the structure of the dispersive coupling. Notice that other schemes based on more complex controls have also been introduced but not implemented as yet \cite{santos2005universal, jacobs2007engineering,strauch2012all}.

Here we propose an alternative approach, for the universal control of single bosonic systems, which is based on the resonant regime offered by the full quantum Rabi model, in this sense extending the original proposal of Ref.~\cite{law1996arbitrary} to systems featuring stronger coupling. The quantum Rabi model~\cite{Rabi:36} has recently attracted renewed attention as it describes light-matter interaction at the most fundamental level. This model, therefore, plays a pivotal role in the quest for quantum-based technologies~\cite{braak2016semi,gu2017microwave}, which has led to theoretical and experimental breakthroughs in the last decade, ranging from its integrability~\cite{Braak:11,Chen:12,xie2017quantum} to the novel phenomena emerging in the ultra- and deep-strong coupling regimes~\cite{Casanova:10,Niemczyk:10,Crespi:12,Yoshihara:17,Langford:17,Forn:17,forn2019ultrastrong}, such as the existence of a quantum phase transition~\cite{Hwang:15,Puebla:17,Cai:21}. Our universal control scheme is gate-based, therefore providing potential for modularity and optimizability, and has the additional feature of naturally achieving some relevant highly populated target states, namely the multi-squeezed states (or generalized squeezed states). The latter, introduced in the 80s~\cite{fisher1984impossibility, hillery1984squeezing, braunstein1987generalized, braunstein1990phase}, have been first realized only recently~\cite{chang2020observation} with the attainment of a multi-squeezed state of the third order (called tri-squeezed hereafter) which, in turn, has been shown to enable universal quantum computation \cite{Zheng:21}. Multi-squeezed states generalize the celebrated standard (second-order) squeezed states to the non-Gaussian regime and in fact, as we will show, they become more and more resourceful for progressively higher orders --- hosting larger levels of Wigner negativities compared to other well known non-Gaussian states. In particular, we will show how to attain tri- and quadri-squeezed states, and provide evidence of the robustness of our generation protocol in the presence of noise for realistic platforms. Moreover, thanks to the ubiquity of the quantum Rabi model in describing the physics of a variety of quantum platforms, the reported method might open new avenues in the design, preparation and control of bosonic states in distinct setups. 


{\em Driven quantum Rabi model.---}
Let us consider a driven two-level system or qubit interacting with a bosonic mode, which can be experimentally realized in a number of quantum platforms, such as in microwave-driven ions~\cite{Woelk:17} or superconducting devices~\cite{Devoret:13,Kjaergaard:19}. The time-dependent Hamiltonian of the system can be written as
\begin{align}\label{eq:Hlab}
H_{\rm lab}= \omega\adaga+g\sigma_x(a+\adag)+H_{\rm d}(t),
  \end{align}
  where $H_{\rm d}(t)=\sum_{j=0,1}\frac{\epsilon_j}{2}\left[\cos(\Delta_j t+\phi_j)\ \sigma_z+\sin (\Delta_j t+\phi_j)\ \sigma_y \right]$ corresponds to the action of two different drivings fields, with frequencies $\Delta_{0,1}$,  amplitudes $\epsilon_{0,1}$ and phases $\phi_{0,1}$. For convenience, we already consider $H_{\rm lab}$ in a rotating frame with respect to the qubit free energy, and assume  $t_0=0$ as initial time, $\sigma_z=\ket{0}\bra{0}-\ket{1}\bra{1}$ and $\sigma^+=\ket{0}\bra{1}$, while the bosonic mode frequency is $\omega$ and its creation and annihilation operators obey $[a,\adag]=1$. Note that the previous Hamiltonian can be transformed to achieve $n$-photon interaction exchange, as well as the standard undriven quantum Rabi model~\cite{Casanova:18,Puebla:19sym}. We now transform $H_{\rm lab}$ with a qubit-dependent displacement operator $T(\alpha)=1/\sqrt{2}(\mathcal{D}^\dagger(\alpha)(\ket{0}\bra{0}-\ket{1}\bra{0})+\mathcal{D}(\alpha)(\ket{0}\bra{1}+\ket{1}\bra{1}))$, with $\mathcal{D}(\alpha)=e^{\alpha a^\dagger-\alpha^*a}$. Up to a constant energy term, the Hamiltonian $H_{\rm a}\equiv T^{\dagger}(-g/\omega) H_{\rm lab} T(-g/\omega)$ reads as
\begin{align}\label{eq:Ha}
H_{\rm a}=\omega\adaga+\sum_{j=0,1}\frac{\epsilon_j}{2}\left[\sigma^+e^{2g(a-\adag)/\omega}e^{-i(\Delta_j t+\phi_j)}+{\rm H.c.} \right].
\end{align}
In the rotating frame of $H_0=\omega\adaga$ with $U_0=e^{-it\omega\adaga}$, $H_{\rm b}=U_0^\dagger (H_{\rm a}-H_{0})U_0$ can be written as
\begin{align}\label{eq:Hb}
H_{\rm b}=\sum_{j=0,1}\frac{\epsilon_j}{2}\left[\sigma^+ e^{2g(a(t)-\adag(t))/\omega}e^{-i(\Delta_jt+\phi_j)}+{\rm H.c.}\right],
  \end{align}
  with $a(t)=ae^{-i\omega t}$. Tuning $\Delta_0=-\Delta_1=- n\omega$ for some $n \in \mathbb{N}$, and requiring that $|\Delta_j|\gg |\epsilon_j|$ together with the Lamb-Dicke condition, $|2g/\omega|\sqrt{\langle (a+\adag)^2\rangle}\ll 1$,  higher order terms in the expansion can be safely neglected. Note however that these requirements may be relaxed depending on the targeted interaction, as we will see later. The previous Hamiltonian $H_{\rm b}$ is then well approximated by $H_{\rm b}\approx H_{\rm n-phot}$, which reads as~\cite{SM}
\begin{align}
H_{\rm n-phot}=\frac{(2g)^n}{2\omega^n \ n!}\left[\sigma^+\left(\epsilon_0  a^n e^{-i\phi_0}+\epsilon_1 (-a^\dagger)^n e^{-i\phi_1}\right)+{\rm H.c.}  \right].\nonumber
\end{align}
 That is, choosing $\phi_0=\phi$, $\phi_1=-\phi+{\rm mod}(n,2)\pi$, and $\epsilon\equiv\epsilon_{0,1}$, it follows
\begin{align}\label{eq:Hnphot}
H_{\rm n-phot}=g_n\sigma_x(a^n e^{-i\phi}+(a^\dagger)^n e^{i\phi}),
\end{align}
with  $g_n=\frac{\epsilon}{2}\frac{(2 g)^n}{\omega^n \ n!}$. Hence, this scheme allows to implement the following family of  gates onto the bosonic degree of freedom,
\begin{align}\label{eq:Gn}
G_{n,\phi}=e^{\mp i\gamma_n(a^n e^{-i\phi}+(\adag)^n)e^{i\phi})},
  \end{align}
assisted by a qubit when prepared in a state $\ket{\pm_x}$ in this transformed frame, such that $\sigma_x\ket{\pm_x}=\pm \ket{\pm_{x}}$, and upon a total evolution time  $t_f=\gamma_n/g_n=\gamma_n 2\omega^n n!/((2g)^n \epsilon)$. This family $G_{n,\phi}$ corresponds to the generalization of squeezing ($n=2$) and displacement ($n=1$) operators \cite{fisher1984impossibility, hillery1984squeezing, braunstein1987generalized, braunstein1990phase}. As we will show later, this scheme naturally yields the generation of multi-squeezed states --- such as tri-squeezed ($n=3$) and quadri-squeezed ($n=4$) states --- and, more in general, can be exploited to engineer other interesting states, such as the cubic-phase state. Before proceeding further, let us remark that such interacting Hamiltonians may be achieved following a different procedure for the particular case of trapped ions in the optical regime, where the Lamb-Dicke parameter allows for a direct coupling of internal states to the $n$th motional sideband~\cite{Leibfried:03,SM}. As we will show in the following, the gates obtained from Eq.~\eqref{eq:Hlab} allow for universal bosonic control.

{\em Universal bosonic control.---} As previously commented, Eq.~\eqref{eq:Gn} includes displacement and squeezing along arbitrary directions ($G_{n,\phi}$ for $n=1$ and $2$, respectively) conditioned to the two-level system state~\cite{SM}. In particular, the squeezing Hamiltonian $H_{\rm 2-phot}=g_2\sigma_x(a^2e^{-i\phi}+{\rm H.c.})$ is obtained for $\epsilon_{0,1}=\epsilon$ and $\phi_0=-\phi_1=\phi$, while the displacement Hamiltonian $H_{\rm 1-phot}=g_1\sigma_x(ae^{-i\phi}+{\rm H.c.})$ for $\phi_1=-\phi+\pi$ and $\phi_0=\phi$. This, together with rotations on the bosonic degree of freedom --- $H_{\rm rot}=-2g_2\sigma_xa^\dagger a$, which can be achieved for $\Delta_0=0$ and $\epsilon_1=0$~\cite{SM} ---  and one of the non-Gaussian gates, $G_{n>2,\phi}$, constitute a universal set of operations for universal bosonic control~\cite{Lloyd:99,Weedbrook:12}. Recall that, on top of rotations, displacement and squeezing, only one non-Gaussian gate is needed to generate any unitary; yet, we stress that our scheme allows for the implementation of different non-Gaussian gates [cf. Eq.~\eqref{eq:Gn} with $n>2$] along different directions in the phase space, by simply tuning the phases and/or frequencies of the driving fields.

\begin{figure}
\includegraphics[width=1\linewidth]{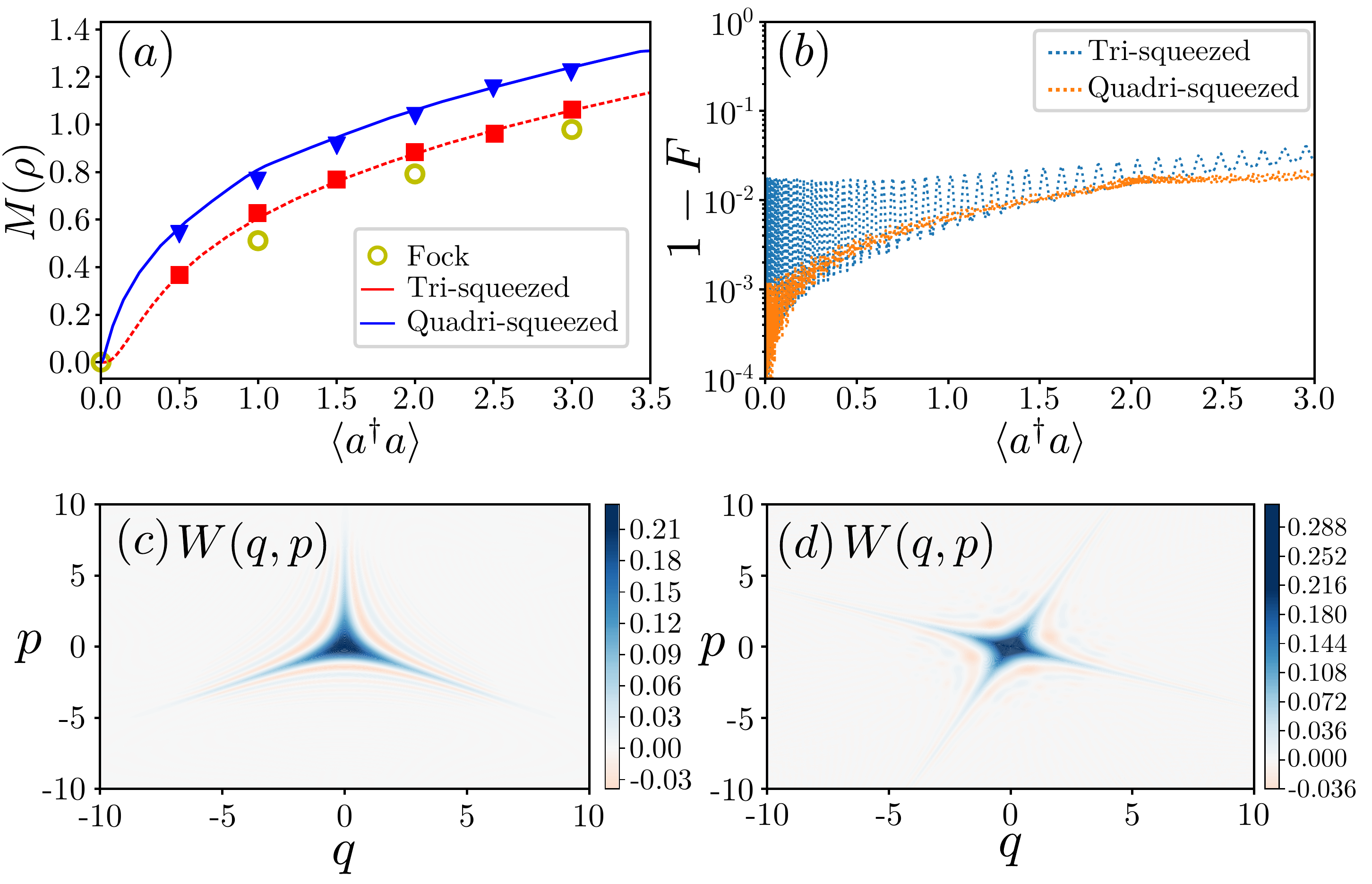}
\caption{\small{(a) Mana of the targeted tri- and quadri-squeezed states $|\gamma_{3,4}\rangle$ (lines)  together with the results following the reported protocol (points) as a function of the energy $\langle a^\dagger a\rangle$. The mana for Fock states is included for reference (open circles). (b) Fidelity between the generated state $\rho$ upon tracing out the two-level system and the targeted tri- and quadri-squeezed, obtained with $g=\omega/10$ and $\epsilon=\omega/5$, and $\omega$, respectively.  Panels (c) and (d) show the Wigner function for the generated tri- and quadri-squeezed states at energy $\langle a^\dagger a \rangle=2$.}}
\label{fig1}
\end{figure}

{\em Multi-squeezed states.---} Let us define multi-squeezed states as $\ket{\gamma_n}=G_{n,0}\ket{0}$ where we fix $\phi=0$ without loss of generality. For $n>2$ these states are of a non-Gaussian nature and hence, they might be a resource enabling universal quantum computation on continuous variables, in addition to Gaussian states and operations \cite{Ferraro05}. As recently shown~\cite{Zheng:21}, the tri-squeezed state $\ket{\gamma_3}$ is one of such resources. Here we show how our scheme allows for a high-fidelity preparation of these non-Gaussian states $\ket{\gamma_{n>2}}$. Besides the fidelity, $F=\bra{\gamma_n}\rho\ket{\gamma_n}$ with $\rho$ the resulting state following our protocol and $\ket{\gamma_n}$ the targeted state, we also characterize them in terms of the Wigner function $W_\rho(q,p)$ in the phase space $q=(a+a^\dagger)/\sqrt{2}$ and $p=i(a^\dagger-a)/\sqrt{2}$. In particular, we compute the Wigner logarithmic negativity or {\em mana}~\cite{Albarelli:18,takagi2018}, defined as $M(\rho)=\log_2\left[\int dqdp |W_\rho(q,p)| \right]$, so that $M(\rho)=0$ for any Gaussian state $\rho$. It is worth mentioning that such multi-squeezed states populate highly excited Fock states even for small values of $\gamma_{n>2}$  and thus, other control strategies may require long sequences to faithfully reproduce these states. Moreover, Fock state populations for these multi-squeezed states decay slower with the increasing Fock number than in standard squeezed states. As an example, the population of Fock states above $m=20$ for tri- and quadri-squeezed states with energy $\langle a^\dagger a\rangle \approx 1$ is non-negligible and amounts to $\sum_{m\geq 20} |\langle m|\gamma_{3,4}\rangle|^2\approx 0.01$. For comparison, the population of these high-excited Fock states for a standard squeezed vacuum state $\ket{\gamma_2}$ with same energy is two orders of magnitude smaller. 

As aforementioned, the generation of multi-squeezed states is achieved in the transformed frame of $H_{\rm n-phot}$ with an initially-prepared qubit state $\ket{\pm_x}$. Yet, it is important to keep in mind the relation between $H_{\rm lab}$ and $H_{\rm n-phot}$ frames, which is given by $\Lambda=T(-g/\omega)U_0$, i.e.  $\ket{\Psi_{\rm lab}(t)}=\Lambda |\Psi_{\rm n-phot}(t)\rangle$. While $U_0$ introduces a phase in the harmonic oscillator state, $T(-g/\omega)$ mixes qubit and bosonic degrees of freedom. This means that an initial state as $|\Psi_{\rm n-phot}(0)\rangle=\ket{\psi_b}\ket{+_x}$ is obtained when  $|\Psi_{\rm lab}(0)\rangle=[|\psi_b^+\rangle\ket{0}+|\psi_b^-\rangle\ket{1}]/2$ with $|\psi_b^{\pm}\rangle=(\mathcal{D}(-g/\omega)\pm \mathcal{D}^\dagger(-g/\omega))\ket{\psi_b}$. The results for the generated multi-squeezed states $|\gamma_{3,4}\rangle$ are shown in Fig.~\ref{fig1}, where we compare the mana $M(\rho)$ of the targeted states $|\gamma_{3,4}\rangle$ with the one obtained under the evolution of Eq.~\eqref{eq:Hlab}, as well as the fidelity with respect the targeted states [cf. Fig.~\ref{fig1}(b)]. The resulting Wigner functions are shown in Fig.~\ref{fig1}(c) and (d), respectively. The previous results showcase the good performance of the method. Note that, in order to retrieve the multi-squeezed state one needs to undo the $\Lambda$ transformation. Yet, since $g/\omega\ll 1$, the initial state can be approximated as $|\Psi_{\rm lab}(0)\rangle\approx \ket{\psi_b}\ket{0}+O(g/\omega)$. Similarly, at the end of the evolution, the state in the $H_{\rm lab}$ frame reads as $|\Psi_{\rm lab}(t_f)\rangle=\Lambda\ket{\gamma_n}\ket{+_x}\approx U_0^\dagger \ket{\gamma_n}\ket{0}+O(g/\omega)$. Such simplification is less demanding for a potential experimental realization at the expense of introducing another source of error. As an example, when approximating $\Lambda$, one obtains a tri-squeezed state for $\langle a^\dagger a\rangle =1$ with a mana $M(\rho)\approx 0.53$ and fidelity $F\approx 0.9$ since $g/\omega=0.1$, as compared to $M(\rho)\approx 0.63$ and $F\approx 0.99$ for an exact $\Lambda$ transformation [cf. Fig.~\ref{fig1}(b)].


In addition, it is worth mentioning that thanks to the mapping between $H_{\rm lab}$ and $H_{\rm n-phot}$ the resulting states can be calculated even without the ability to perform the transformation $\Lambda$ (see~\cite{SM} for details). Indeed, starting with $\ket{\Psi_{\rm lab}(0)}=\ket{\psi_b}\ket{+_x}$, the resulting state acquires a cat-like shape. Upon the evolution, the state conditioned to the qubit in $\ket{\pm_x}$ becomes $\ket{\Psi_{\rm lab}(t)}\propto (G_{n,\phi}\pm G_{n,\phi+\pi})\ket{\psi_b}\ket{\pm_x}$. This corresponds to standard cat states for $n=1$, and a cat-like superposition of squeezed states for $n=2$. For $n>2$ this generalizes to cat-like superpositions of multi-squeezed states. Such cat states can feature a large negativity or mana for a fixed energy~\cite{SM}. 


{\em Cubic-phase gate.---} From the family of non-Gaussian unitaries $G_{n,\phi}$~ in Eq.\eqref{eq:Gn} one can generate a cubic-phase gate, 
\begin{align}\label{eq:Gc}
G_{c}=e^{-i\gamma_{c}(a+\adag)^3},
\end{align}
with {\em cubicity} $\gamma_c$. 
Such gate can be achieved exploiting a non-Gaussian interaction since $[H_{1s},[H_4,H_{2s}]]=8 g_1 g_2 g_4 (a+a^\dagger)^3$ where $H_{n}=g_n (a^n+(a^\dagger)^n)$ and $H_{ns}=ig_n((a^{\dagger})^n-a^n)$. Such relation motivates the use of a Trotter evolution, since $e^{iAt}e^{iBt}e^{-iAt}e^{-iBt}=e^{-[A,B]t^2}+O(t^3)$~\cite{Lloyd:99} for any operator $A$ and $B$. Combining these two relations, we find~\cite{SM}
\begin{align}\label{eq:Ublock}
    U_{\rm block}=U_1 U_{4}^\dagger U_2^\dagger U_4 U_2 U_1^\dagger U_2^\dagger U_{4}^\dagger U_2 U_4\approx e^{-i\delta (a+a^\dagger)^3}
\end{align}
with $U_{1}=e^{-i \tau_1 H_{1s}}$, $U_2=e^{-i \tau_2 H_{2s}}$ and $U_4=e^{-i\tau_4 H_4}$, and $\delta=8\tau_1g_1\tau_2g_2\tau_4g_4\ll 1$, being the times $\tau_{1,2,4}$ such that $U_{0}=\mathbb{I}$. The unitaries $U_{1,2,4}$ follow directly from Eq.~\eqref{eq:Hnphot} with $\phi=\pi/2$, $\pi/2$ and $0$, respectively, assisted by the two-level state in $\ket{+_x}$. To the contrary, $U_{1,2,4}^\dagger$ are achieved setting either $\phi\rightarrow\phi+\pi$ or upon a $\pi$-pulse of the two-level system, $\ket{+_x}\rightarrow \ket{-_x}$.  Recall, however, that the rotation must be realized in the frame of $H_{\rm n-phot}$, and therefore, such operation must be appropriately translated into the $H_{\rm lab}$ frame. In particular, a rotation around the $z$ axis for $H_{\rm n-phot}$  corresponds to a rotation around the $x$ axis in the $H_{\rm lab}$ frame~\cite{SM}. Repeating $N$ times the sequence in Eq.~\eqref{eq:Ublock}, we arrive to $(U_{\rm block})^N\approx e^{-i\delta N (a+a^\dagger)^3}$ which leads to $G_c$ with a cubicity $\gamma_c=N\delta$. 

In order to meet the conditions to achieve Eq.~\eqref{eq:Ublock}, small $g/\omega$ and $\epsilon/\omega$ values are required, so that the errors do not accumulate and do not spoil the approximate mapping between $H_{\rm lab}$ and $H_{\rm n-phot}$ as well as the Trotterization. This, in turn, leads to long evolution times to obtain a significant cubicity $\gamma_c$. As proof-of-principle, we aim at reaching $\gamma_c=1/\sqrt{32\pi}\approx 0.1$, as required to perform a $T$ gate on GKP encoded qubits~\cite{Douce:19}. Choosing $g=\omega/20$ and $\epsilon=\omega/100$, with $\tau_1=\tau_2/2=\tau_4/80=100\omega$, we find $\delta\approx 8.27\cdot 10^{-5}$. Upon $N=1200$ repetitions of the block~\eqref{eq:Ublock}, we obtain $\gamma_c\approx 0.1$. Under these parameters, a cubic-phase state, $\ket{\gamma_c}=G_c\ket{0}$, is prepared with a fidelity $F\approx 0.99$, whose mana matches the targeted one, i.e. $M(\rho)\approx 0.14$. This method, however, may demand long evolution times to generate $G_c$, possibly longer than typical coherence times in state-of-the art quantum platforms~\footnote{It is important to remark that the parameters $\epsilon$, $\tau_{1,2,4}$ and $g$ can be optimized to reduce the duration of the sequence}, and thus other approximate schemes may be better placed for their experimental implementation (see for example Ref.~\cite{Zheng:21}).  Yet, this theoretical example is a good illustration of the universal bosonic control granted by $H_{\rm lab}$.


\begin{figure}
\centering
\includegraphics[width=1\linewidth]{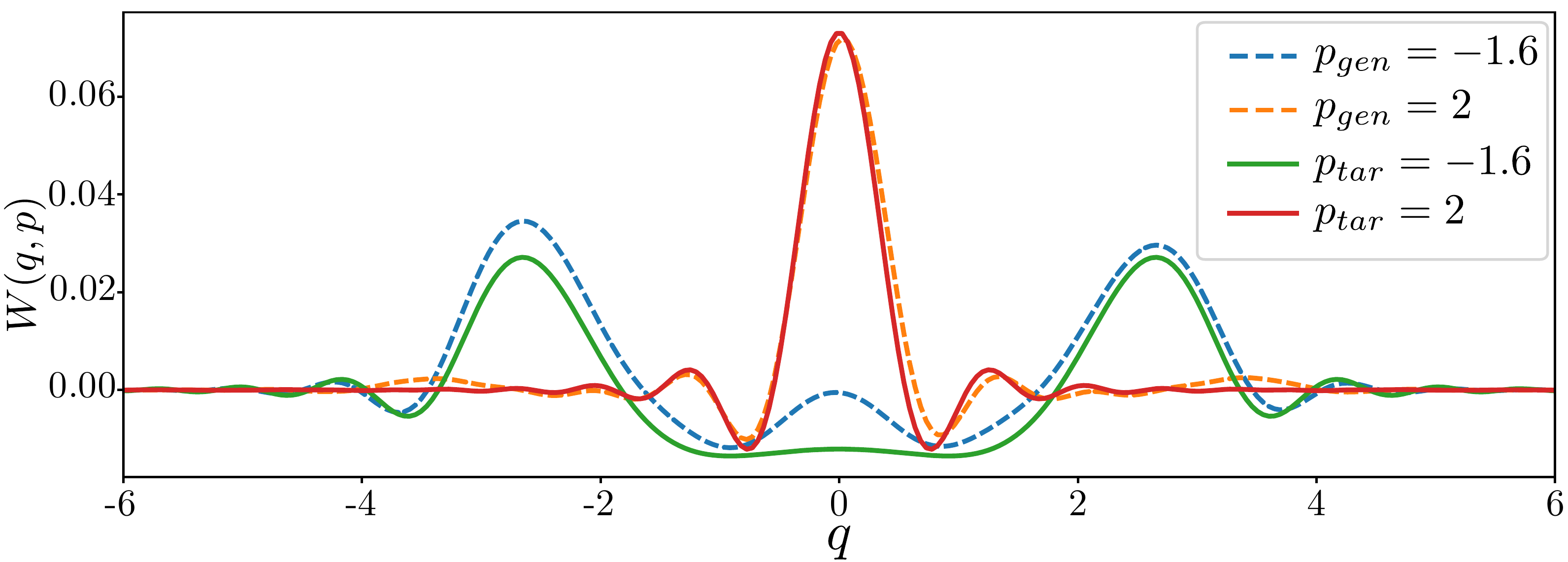}
\caption{\small{Cross sections of the Wigner function $W(q,p)$ for $p=-1.6$ and $p=2$, for an ideal tri-squeezed state $|\gamma_3^m\rangle$ with $\gamma_3^m\approx 0.116$ (solid lines) and the generated state $\rho$ under Eq.~\eqref{eq:rhoME} with $g=\omega/20$ and $\epsilon=\omega$ (dashed lines), with a fidelity $F\approx 0.94$. The ideal and generated mana are $M\approx 0.22$ and $0.14$, respectively. See main text for further details.}}
\label{fig2}
\end{figure}

{\em Decoherence effects and experimental feasibility.---} The Hamiltonian considered in Eq.~\eqref{eq:Hlab} describes a variety of quantum setups where a qubit is coupled to a bosonic degree of freedom, such as in microwave-driven trapped ions~\cite{Mintert:01,Timoney:11,Weidt:16,Piltz:16} or superconducting qubit devices~\cite{Devoret:13,Kjaergaard:19}. Hence, the reported scheme can in principle be used to generate multi-squeezed states in these setups. Yet, the impact of decoherence may hinder the overall performance and quality of the desired operations. In order to address the experimental feasibility, we focus on a platform inspired from Ref.~\cite{Khosla:18} and consider a qubit coupled to a nanomechanical resonator. The qubit consists of a pair of superconducting electrodes coupled through SQUID junctions, which also allows for a tunable qubit-nanomechanical resonator coupling. The Hamiltonian can be written as~\cite{Khosla:18,Armour:02} $H=\nu\sigma_z/2+\omega a^\dagger a+g \sigma_z (a+a^\dagger)$, which together with the application of qubit drivings and a qubit rotation corresponds to Eq.~\eqref{eq:Hlab}. Considering $\omega= 2\pi\times 200$ MHz, $g= 2\pi \times 10$ MHz and amplitudes of the driving fields $\epsilon=\omega$ we find that the required time to prepare tri- and quadri-squeezed states $|\gamma_{3,4}\rangle$ would result in $\tau_3\approx \gamma_3\cdot 9.5\ \mu$s, and $\tau_4\approx \gamma_4\cdot 380 \ \mu$s. In addition, we consider an optimistic qubit coherence time of $T_2\approx 1$ms~\cite{Kjaergaard:19}, together with a mechanical resonator relaxation rate $\kappa\approx 2\pi \times 7$ kHz~\cite{Khosla:18}. The impact of decoherence effects is modelled by a Lindblad master equation~\cite{Breuer}, 
\begin{align}\label{eq:rhoME}
    \dot{\rho}=-i[H_{\rm lab},\rho]+\mathcal{D}_a[\rho]+\mathcal{D}_{a^\dagger}[\rho]+\mathcal{D}_{\sigma_z}[\rho]+\mathcal{D}_{\sigma^-}[\rho],
\end{align}
with $\mathcal{D}_A[\rho]=\Gamma_A(A\rho A^\dagger-\{A^\dagger A,\rho \}/2)$ the dissipator for a jump operator $A$, so that $\Gamma_a=\Gamma_{a^\dagger}=3.5\cdot 10^{-5}\omega$,  $\Gamma_{\sigma_z}=2\Gamma_{\sigma^-}=5\cdot 10^{-6}\omega$. Under this noisy evolution, we aim at preparing $|\gamma_{3,4}\rangle$. In contrast to the fully-coherent scenario, the generated amplitudes $\gamma_{3,4}$ are reduced by decoherence effects. For this reason, we computed the value of the amplitudes $\gamma_{3,4}^m$ that maximize the fidelity between the state after a time $t$ and a state $|\gamma_{3,4}\rangle$ --- \textit{e.g.}, $F=\max_{\gamma_3}\langle \gamma_3 |\rho |\gamma_3\rangle=\langle \gamma_3^m |\rho |\gamma_3^m\rangle$ for a tri-squeezed state.  

In Fig.~\ref{fig2} we show an example for the cross-sections of the Wigner function corresponding to the generated tri-squeezed state under Eq.~\eqref{eq:rhoME}, compared to the ideal state, which clearly reveal negative regions. The fidelity amounts to $F\approx 0.94$ for $\gamma_3^m\approx 0.116$ and the generated state reveals a significant amount of mana $M(\rho)\approx 0.14$, even if smaller than the ideal value of $M(|\gamma_3^m\rangle\langle \gamma_3^m|)=0.22$. Recall that we use $\Delta_{0,1}=\mp 3\omega$, $\phi_{0}=0$, $\phi_1=\pi$ to generate a tri-squeezed state, while we set $\gamma_3=0.2$ so that the evolution time amounts to $\tau_3=1.9 \ \mu$s. Regarding the generation of a quadri-squeezed state ($\Delta_{0,1}=\mp 4\omega$, $\phi_{0,1}=0$), a coupling $g=\omega/20$ demands very long evolution times, and thus the state is eventually spoiled by distinct decoherence sources. For example, setting $\gamma_4=0.05$ results in $\tau_4=19\ \mu$s and in a generated state with small negativity and an amplitude $\gamma_4^{m}\ll 1$. However, notice that the quality of the preparation can be significantly enhanced in platforms that have access to ultrastrong qubit-boson coupling regimes~\cite{Kockum:19}. Indeed, under the same noisy evolution, increasing the coupling to $g=\omega/10$, we obtain fidelities $F\approx 0.95$ and $0.96$ for tri- and quadri-squeezed states with amplitudes $\gamma_3^m\approx 0.131$ and $\gamma_4^m\approx 0.025$ (setting $\gamma_3=0.2$ and $\gamma_4=0.0375$) and whose mana is close to the target, $M(\rho)\approx 0.31$ instead of $M(|\gamma_3^m\rangle\langle \gamma_3^m|)=0.33$, and $M(\rho)\approx 0.12$ instead of $M(|\gamma_4^m\rangle\langle \gamma_4^m|)=0.15$, respectively.

{\em Conclusions.---} As we have shown, a simple qubit-bosonic mode system plus driving fields acting solely on the qubit degree of freedom can be used to prepare the bosonic mode in multi-squeezed states, such as tri-squeezed and quadri-squeezed states, as well as cat-like superpositions thereof.  Moreover, thanks to the possibility to generate non-Gaussian operations, this model provides indeed universal bosonic control. As a proof-of-principle, we illustrated how this system can be used to generate cubic-phase gates. Finally, we showed the impact of distinct decoherence sources in the reported scheme focusing on a particular experimental platform. As expected, decoherence spoils the good performance of the method but still allows to obtain multi-squeezed states with high fidelity displaying significant negativities in the phase space. Thanks to the ubiquity of the considered driven quantum Rabi model, our results could open the door for the preparation of resourceful bosonic states in a variety of quantum platforms.



%


\widetext
\clearpage
\begin{center}
\textbf{\large Supplemental Material}\\ \vspace{0.2cm} \textbf{\large Multi-squeezed state generation and universal bosonic control via a driven quantum Rabi model}
\end{center}
\setcounter{equation}{0}
\setcounter{figure}{0}
\setcounter{table}{0}
\setcounter{section}{0}
\setcounter{page}{1}
\makeatletter
\renewcommand{\theequation}{S\arabic{equation}}
\renewcommand{\thefigure}{S\arabic{figure}}
\renewcommand{\bibnumfmt}[1]{[S#1]}
\renewcommand{\citenumfont}[1]{S#1}

\begin{center}
Peter McConnell${}^1$, Alessandro Ferraro${}^{1,2}$, Ricardo Puebla${}^{3,4}$\\ \vspace{0.2cm}
\textit{\small{${}^1$Centre for Theoretical Atomic, Molecular and Optical Physics, Queen's University Belfast, Belfast BT7 1NN, United Kingdom}}\\

\textit{\small{${}^2$Quantum Technology Lab, Dipartimento di Fisica Aldo Pontremoli, Universit\`a degli Studi di Milano, I-20133 Milano, Italy}}\\

\textit{\small{${}^3$Instituto de F{\'i}sica Fundamental (IFF), CSIC, Calle Serrano 113b, 28006 Madrid, Spain}}\\

\textit{\small{${}^4$Departamento de F{\'i}sica, Universidad Carlos III de Madrid, Avda. de la Universidad 30, 28911 Legan{\'e}s, Spain}}
\end{center}

\section{Multi-photon interaction terms from a driven quantum Rabi model: From $H_{\rm lab}$ to $H_{\rm n-phot}$}\label{s:theory}
Let us consider a qubit or a two-level system subject to two drivings, and coupled to a bosonic mode. The time-dependent Hamiltonian of the system can be written as
\begin{align}\label{eqSM:Hlab}
H_{\rm lab}= \omega\adaga+g\sigma_x(a+\adag)+\sum_{j=0}^{1}\frac{\epsilon_j}{2}\left[\cos(\Delta_j t+\phi_j)\ \sigma_z+\sin (\Delta_j t+\phi_j)\ \sigma_y \right].
  \end{align}
The drivings have frequency $\Delta_j$, amplitude $\epsilon_j$ and a phase $\phi_j$. We will refer to the evolution under $H_{\rm lab}$ as the {\em laboratory} framework. Note that we assume $t_0=0$, $\sigma_z=\ket{0}\bra{0}-\ket{1}\bra{1}$, and $\sigma^+=\ket{0}\bra{1}$. We now transform $H_{\rm lab}$ with a qubit-dependent displacement operator $T(\alpha)$ defined as (in the qubit basis)
\begin{align}
T(\alpha)=\frac{1}{\sqrt{2}}\begin{bmatrix}\mathcal{D}^{\dagger}(\alpha) & \mathcal{D}(\alpha) \\ -\mathcal{D}^{\dagger}(\alpha) & \mathcal{D}(\alpha) \end{bmatrix} \quad {\rm with} \quad \mathcal{D}(\alpha)=e^{\alpha \adag-\alpha^*a}.
  \end{align}
For convenience, let's write the following relations
\begin{align}
  T^{\dagger}(\alpha)\adaga T(\alpha)&=\adaga+|\alpha|^2-\sigma_z(a \alpha^*+\adag\alpha),\\
  T^{\dagger}(\alpha)\sigma_xT(\alpha)&= - \sigma_z,\\
  T^{\dagger}(\alpha)\sigma_y T(\alpha)&= -i\mathcal{D}(2\alpha)\sigma^++{\rm H.c.},\\
  T^{\dagger}(\alpha)\sigma_z T(\alpha)&= \mathcal{D}(2\alpha)\sigma^++{\rm H.c.},\\
  T^{\dagger}(\alpha)\sigma_x(a+\adag) T(\alpha)&=-\sigma_z(a+\adag)+2{\rm Re}[\alpha].
\end{align}
Choosing a displacement amplitude of $-g/\omega$ the linear spin-boson coupling is removed. The transformed Hamiltonian reads as (up to a constant energy shift)
\begin{align}
H_{\rm a}\equiv T^{\dagger}(-g/\omega) H_{\rm lab} T(-g/\omega)= \omega\adaga+\sum_{j=0}^{n_d}\frac{\epsilon_j}{2}\left[\sigma^+e^{2g(a-\adag)/\omega}e^{-i(\Delta_j t+\phi_j)}+{\rm H.c.} \right].
\end{align}
Finally, moving to an interaction picture with respect to $H_0=\omega\adaga$ with $U_0=e^{-it\omega\adaga}$, we obtain
\begin{align}\label{eqSM:Hb}
H_{\rm b}\equiv U_0^{\dagger}(H_{\rm a}-H_0)U_0=\sum_{j=0}^{n_d}\frac{\epsilon_j}{2}\left[\sigma^+ e^{2g(a(t)-\adag(t))/\omega}e^{-i(\Delta_jt+\phi_j)}+{\rm H.c.}\right].
  \end{align}
Hence, tuning the frequencies $\Delta_j=\pm \omega n$, and requiring  $|\Delta_j|\gg |\epsilon_j|$ (to perform a RWA) together with the Lamb-Dicke regime, $|2g/\omega|\sqrt{\langle (a+\adag)^2\rangle}\ll 1$ (to safely neglect higher order terms in the expansion of the exponential), the previous Hamiltonian $H_{\rm c}$ can be well approximated by
\begin{align}\label{eqSM:ctonphot}
  H_{\rm b}\approx H_{\rm n-phot}=\sum\limits_{\substack{k \\\Delta_k=-k\omega}}\frac{\epsilon_k}{2}\frac{(2g)^k}{\omega^k\ k!}\left[\sigma^+a^ke^{-i\phi_k}+{\rm H.c.}\right]+\sum\limits_{\substack{k \\ \Delta_k=k\omega}}\frac{\epsilon_k}{2}\frac{(2g)^k}{\omega^k\ k!}\left[\sigma^+(-\adag)^k e^{-i\phi_k}+{\rm H.c.}\right].
  \end{align}
To generate an interaction of the type $\sigma_x(a^n+(\adag)^n)$ one needs $\Delta_0=-n\omega$ and $\Delta_1=n\omega$. Choosing $\phi_0=0$ and $\phi_1={\rm mod}(n,2)\pi$, and $\epsilon_{0}=\epsilon_1$, we find
\begin{align}
H_{\rm n-phot}=g_n\sigma_x(a^n+(\adag)^n), \quad {\rm with} \quad g_n=\frac{\epsilon_0}{2}\frac{(2 g)^n}{\omega^n \ n!}.
\end{align}
Therefore, in order to generate the unitary
\begin{align}
G_k=e^{-i\gamma_k(a^k+(\adag)^k)},
  \end{align}
the total evolution time amounts to $t_f=\gamma_k/g_k=\gamma_k 2\omega^k k!/((2g)^k \epsilon_0)$. In a similar manner, one can generate interactions $a^{\dagger,n}-a^n$. For that, however, the phases (or $\epsilon$) have to be tuned differently. Indeed, the Hamiltonian
\begin{align}
H_{\rm n-phot;s}=ig_n \sigma_y(a^{\dagger,n}-a^n),
  \end{align}
is attained again by tuning $\Delta_0=-n\omega$ and $\Delta_1=n\omega$, and either (i) $\epsilon=\epsilon_{0}=\epsilon_1$ and $\phi_0=\pi$, $\phi_1={\rm mod}(n,2)\pi$, or (ii) $\epsilon=-\epsilon_0=\epsilon_1$ and $\phi_0=0$, $\phi_1={\rm mod}(n,2)\pi$.

\subsection{Optical ions}
For the specific setup of optical trapped ions the procedure can be simplified as $n$-sidebands can be directly driven~\cite{Leibfried:03,Haffner:08}. The Hamiltonian reads as
\begin{align}
H_{\rm oi}=\frac{\omega_I}{2}\sigma_z+\nu \adaga+\sum_j \Omega_j\sigma_x\cos(\omega_jt-k_jx+\phi_j),
  \end{align}
where the wavevector $k$ is such that $\eta_j=k_jx_0=k_j(2m\nu)^{-1/2}$, being $\eta_j$ the  Lamb-Dicke parameter, $m$ the mass of the ion and $x=x_0(\adag+a)$ and $x_0$ the zero-point motion of the ion.  Then, by moving to an interaction picture with respect to $H_0=(\omega_I/2)\sigma_z+\nu\adaga$, and applying the optical RWA since $|\omega_I+\omega_j|\gg |\omega_I-\omega_j|,\Omega_j,\nu$, we get
\begin{align}\label{eq0:4sim}
H^I=\sum_j \frac{\Omega_j}{2}\left[\sigma^+ e^{i(\omega_I-\omega_j)t}e^{i\eta_j(a(t)+\adag(t))}e^{-i\phi_j}+{\rm H.c.}\right],
  \end{align}
where $a(t)=ae^{-i\nu t}$.  Now one can driven $n$-photon sidebands by tuning $\Delta =\omega_I-\omega_j=\pm n\nu$, followed by Lamb-Dicke regime plus RWA. Choosing the red-sideband $\Delta_0=\omega_I-\omega_0=+n\nu$, we obtain
\begin{align}
H^I=\frac{\Omega_0}{2}\left[\frac{(i\eta_0)^n}{n!}\sigma^+a^ne^{-i\phi_0}+{\rm H.c}\right].
  \end{align}
In this manner, by selecting $\Delta_{0,1}=\pm n\nu$, $\eta\equiv\eta_{0,1}$, $\Omega\equiv \Omega_{0,1}$ and $\phi_{0,1}=n\pi/2$, one finds 
\begin{align}
H^I=g_n\sigma_x(a^n+(\adag)^n) \qquad  {\rm with}\qquad g_n=\frac{\Omega \eta^n}{2\  n!}.
\end{align}



\section{Displacement, squeezing and rotations}
Considering $\epsilon_{0,1}=\epsilon$ and $\Delta_{0}=-\omega$, $\Delta_{1}=+\omega$, Eq.~\eqref{eqSM:ctonphot} reduces to
\begin{align}
    H_{\rm 1-phot}&=g_1\left[\sigma^+ae^{-i\phi_0}-\sigma^+a^\dagger e^{-i\phi_1}+{\rm H.c.}\right]=g_1\sigma_x\left(ae^{-i\phi}+a^\dagger e^{i\phi}\right)
\end{align}
where in the last step we have used $\phi_0=\phi$ and $\phi_1=-\phi+\pi$. Preparing the qubit in a $\ket{\pm_x}$ state, we implement the following unitary on the bosonic degree of freedom upon a evolution time $t$,
\begin{align}
    U=e^{\mp ig_1 t( ae^{-i\phi}+a^\dagger e^{i\phi})}=\mathcal{D}[\alpha]
\end{align}
where $\mathcal{D}[\alpha]=e^{\alpha a^\dagger-\alpha^* a}$ is a displacement operator, whose amplitude takes the value $\alpha=\mp i g_1 t e^{i\phi}$.  Similarly, setting $\epsilon_{0,1}=\epsilon$, $\Delta_{0}=-2\omega$, $\Delta_{1}=+2\omega$ and $\phi_{0}=-\phi_1=\phi$, Eq.~\eqref{eqSM:ctonphot} reduces to
\begin{align}
H_{\rm 2-phot}=g_2\sigma_x\left(a^2 e^{-i\phi}+(a^\dagger)^2e^{i\phi}\right),
\end{align}
and again, the evolution conditioned to the qubit prepared in a state $\ket{\pm_x}$ leads to
\begin{align}
    U=e^{\mp i g_2 t (a^2 e^{-i\phi}+(a^\dagger)^2e^{i\phi})}=\mathcal{S}[z],
\end{align}
where $\mathcal{S}[r]=e^{1/2(r^* a^2-r(a^\dagger)^2)}$ is a squeezing operator, with parameter $r=\pm 2i g_2 te^{i\phi}$.  In both cases, i.e. for  displacement and squeezing, the direction can be controlled by tuning the phase $\phi$. 

In addition, one can generate rotations on the bosonic degree freedom following the same scheme. For that, we turn our attention to Eq.~\eqref{eqSM:Hb}. Let us consider only one driving with zero frequency, namely, $\epsilon_1=0$ and $\epsilon_0=\epsilon$ with $\Delta_0=0$ and $\phi_0=0$ for convenience.  Then, Eq.~\eqref{eqSM:Hb} can be expanded up to second order $2g/\omega$ as
\begin{align}
    H_{\rm b}\approx \frac{\epsilon}{2}\sigma^+\left[\mathds{1}+\frac{2g}{\omega}\left(ae^{-i\omega t}+a^\dagger e^{i\omega t}\right)+\frac{4g^2}{2\omega^2}\left(a^2e^{-2i\omega t}+(a^\dagger)^2e^{2i\omega t}-2a^\dagger a-\mathds{1}\right)\right]+{\rm H.c.}.
\end{align}
Now, as considered before $\omega\gg \epsilon$, which together $|2g/\omega|\sqrt{\langle(a+a^\dagger)^2 \rangle}\ll 1$, allows us to perform a RWA,
\begin{align}
    H_{\rm b}\approx (g_0-g_2)\sigma_x-2g_2\sigma_x a^\dagger a. 
\end{align}
The evolution under the previous Hamiltonian leads to a rotation by an angle $\theta=\mp 2 g_2 t$, i.e. $R(\theta)=e^{i\theta a^\dagger a}$, provided the qubit is in a $\ket{\pm_x}$ state. 

\section{Relation between $H_{\rm lab}$ and $H_{\rm n-phot}$ frames}
Following the transformation required to bring $H_{\rm lab}$ in the form of $H_{\rm n-phot}$ we find that the states are related as
\begin{align}
    |\Psi_{\rm lab}(t)\rangle = \Lambda |\Psi_{\rm n-phot}(t)\rangle
\end{align}
where $\Lambda=T(-g/\omega)U_0$. This affects the initial and final states in the {\em lab} frame, as well as the operations that need to be performed in the {\em n-phot} frame. For example, in order to perform a qubit rotation around the $z$-axis in the {\em n-phot} frame,  $R_{z;\theta}=e^{-i\theta \sigma_z}$, one needs to to perform $T(-g/\omega)R_{z;\theta}T^{\dagger}(-g/\omega)=R_{x,-\theta}=e^{-i(-\theta)\sigma_x}$ in the {\em lab} frame.

Considering an arbitrary spin state $\ket{\varphi_s}=(|\alpha|^2+|\beta|^2)^{-1/2}(\alpha\ket{0}+\beta\ket{1})$, so that $|\Psi_{\rm n-phot}\rangle=\ket{\varphi_b}\ket{\varphi_s}$, then 
\begin{align}\label{eqSM:psilab}
|\Psi_{\rm lab}\rangle= \frac{1}{\sqrt{2(|\alpha|^2+|\beta|^2)}}\left[\ket{0}|\varphi_b^+\rangle+\ket{1}|\varphi_b^-\rangle\right],
  \end{align}
where we have neglected the phase introduced by $U_0=e^{-i\omega t a^\dagger a}$, assuming, for example that $\omega t=2\pi n$ with $n\in \mathbb{N}$, and with $|\varphi_b^{\pm}\rangle=[\beta \mathcal{D}(-g/\omega)\pm \alpha \mathcal{D}^\dagger(-g/\omega)]\ket{\varphi_b}$.

Measuring the qubit state in the $\sigma_x$ basis, in the {\em lab} frame, the state in Eq.~\eqref{eqSM:psilab} is projected to 
\begin{align}
  \frac{\Pi_\pm |\Psi_{\rm lab}\rangle}{\sqrt{\langle \Psi_{\rm lab}| \Pi_{\pm}^{\dagger} \Pi_{\pm}  |\Psi_{\rm lab}\rangle}}= 2\mathcal{D}(\mp g/\omega)\ket{\varphi_b}\ket{\pm_x}
\end{align}
where $\Pi_{\pm}$ stands for the projector onto $\ket{\pm_x}$.

Let us assume that the initial state in the {\em lab} frame is given by $|\Psi_{\rm lab}(0)\rangle=\ket{\varphi_b}\ket{\varphi_s}$, with again a generic qubit state $\ket{\varphi_s}=(|\alpha|^2+|\beta|^2)^{-1/2}(\alpha\ket{0}+\beta\ket{1})$. Then, the initial state in the {\em n-phot} frame is
\begin{align}
    |\Psi_{\rm n-phot}(0)\rangle=T^\dagger(-g/\omega) |\Psi_{\rm lab}(0)\rangle= \frac{1}{\sqrt{2(|\alpha|^2+|\beta|^2)}}\left[(\alpha-\beta)\mathcal{D}(-g/\omega)\ket{\varphi_b}\ket{0}+(\alpha+\beta)\mathcal{D}^\dagger(-g/\omega)\ket{\varphi_b}\ket{1} \right]
\end{align}
Upon the application of a Hamiltonian $H_{\rm n-phot}=g_n \sigma_x(a^n+(a^\dagger)^n)$, the evolved state reads as
\begin{align}
    |\Psi_{\rm n-phot}(t)\rangle=&\frac{1}{2\sqrt{|\alpha|^2+|\beta|^2}}\left[\ket{+_x}e^{-ig_nt(a^n+(a^\dagger)^n)}\left[(\alpha-\beta)\mathcal{D}(-g/\omega)+(\alpha+\beta)\mathcal{D}^\dagger(-g/\omega) \right]\ket{\varphi_b}\right]\\ &\left.+\ket{-_x}e^{ig_n t(a^n+(a^\dagger)^n)}\left[ (\alpha-\beta)\mathcal{D}(-g/\omega)-(\alpha+\beta)\mathcal{D}^\dagger(-g/\omega)\right]|\varphi_b\rangle\right]
\end{align}
In the {\em lab} frame, such state reads as (assuming $U_0=\mathbb{I}$ for convenience)
\begin{align}
    |\Psi_{\rm lab}(t)\rangle=& \frac{1}{4\sqrt{|\alpha|^2+|\beta|^2}}\left[ \ket{0}\left(\mathcal{D}^\dagger(-g/\omega) \left\{ e^{-it g_n(a^n+(a^\dagger)^n)}[(\alpha-\beta)\mathcal{D}(-g/\omega)+(\alpha+\beta)\mathcal{D}^\dagger(-g/\omega)]\ket{\varphi_b}\right.\right.\right.\\&\left.\left.\left.+e^{itg_n(a^n+(a^\dagger)^n)}[(\alpha-\beta)\mathcal{D}(-g/\omega)-(\alpha+\beta)\mathcal{D}^\dagger(-g/\omega)]\ket{\varphi_b}\right\}\right.\right.\\&+\mathcal{D}(-g/\omega) \left\{ e^{-it g_n(a^n+(a^\dagger)^n)}[(\alpha-\beta)\mathcal{D}(-g/\omega)+(\alpha+\beta)\mathcal{D}^\dagger(-g/\omega)]\ket{\varphi_b}\right\}\\&\left.\left.-e^{itg_n(a^n+(a^\dagger)^n)}[(\alpha-\beta)\mathcal{D}(-g/\omega)-(\alpha+\beta)\mathcal{D}^\dagger(-g/\omega)]\ket{\varphi_b}\right\}\right)\\&+\ket{1}\left(-\mathcal{D}^\dagger(-g/\omega) \left\{ e^{-it g_n(a^n+(a^\dagger)^n)}[(\alpha-\beta)\mathcal{D}(-g/\omega)+(\alpha+\beta)\mathcal{D}^\dagger(-g/\omega)]\ket{\varphi_b}\right.\right.\\&\left.\left.\left.+e^{itg_n(a^n+(a^\dagger)^n)}[(\alpha-\beta)\mathcal{D}(-g/\omega)-(\alpha+\beta)\mathcal{D}^\dagger(-g/\omega)]\ket{\varphi_b}\right\}\right.\right.\\&+\mathcal{D}(-g/\omega) \left\{ e^{-it g_n(a^n+(a^\dagger)^n)}[(\alpha-\beta)\mathcal{D}(-g/\omega)+(\alpha+\beta)\mathcal{D}^\dagger(-g/\omega)]\ket{\varphi_b}\right\}\\&\left.\left.-e^{itg_n(a^n+(a^\dagger)^n)}[(\alpha-\beta)\mathcal{D}(-g/\omega)-(\alpha+\beta)\mathcal{D}^\dagger(-g/\omega)]\ket{\varphi_b}\right\}\right).
\end{align}
For $\alpha=\beta$, this simplifies to
\begin{align}\label{eqSM:psilabCat}
|\Psi_{\rm lab}(t)\rangle=& \frac{1}{2\sqrt{2}}\left[\ket{0}\left(\mathcal{D}^\dagger G_n \mathcal{D}^\dagger -\mathcal{D}^\dagger G_n^\dagger \mathcal{D}^\dagger+\mathcal{D}G_n\mathcal{D}^\dagger+\mathcal{D}G_n^\dagger \mathcal{D}^\dagger\right)\ket{\varphi_b}\right.\nonumber \\&+\left.\ket{1}\left(-\mathcal{D}^\dagger G_n\mathcal{D}^\dagger+\mathcal{D}^\dagger G_n^\dagger \mathcal{D}^\dagger+\mathcal{D}G_n\mathcal{D}^\dagger+\mathcal{D}G_n^\dagger \mathcal{D}^\dagger \right)\ket{\varphi_b}\right]
\end{align}
with $\mathcal{D}=\mathcal{D}(-g/\omega)$ and $G_n=e^{-itg_n(a^n+(a^\dagger)^n)}$. Approximating $\mathcal{D}\approx \mathbb{I}+O(g/\omega)$, we find 
\begin{align}
    |\Psi_{\rm lab}(t)\rangle\approx \frac{1}{\sqrt{2}}\left[\ket{0}G_n\ket{\varphi_b}+\ket{1}G_n^\dagger\ket{\varphi_b}\right].
\end{align}
Projecting the qubit onto $\ket{+_x}$ we arrive to $|\Psi_{\rm lab}\rangle\propto \ket{+_x}\left(\mathcal{D}G_n\mathcal{D}^\dagger+\mathcal{D}G_n^\dagger \mathcal{D}^\dagger\right)\ket{\varphi_b}=\ket{+_x}\mathcal{D}\left(G_n+G_n^\dagger\right)\mathcal{D}^\dagger\ket{\varphi_b}$. Similarly, finding the qubit  in the $\ket{0}$ state, it follows $|\Psi_{\rm lab}(t)\rangle\propto \ket{0}\left[\mathcal{D}^\dagger(G_n-G_n^\dagger)+\mathcal{D}(G_n+G_n^\dagger)\right]\mathcal{D}^\dagger \ket{\varphi_b}$.

\section{Cat-like superposition of multi-squeezed states}
Following the derivation of the previous Section, an initial state $\ket{\Psi_{\rm lab}(0)}=\ket{+_x}\ket{0}$ results in $\ket{\Psi_{\rm lab}(t)}$ given in Eq.~\eqref{eqSM:psilabCat}. Upon a projection of the qubit onto $\ket{+_x}$, the bosonic degree of freedom is left in the cat-like superposition $|\Psi_{\rm lab}(t)\rangle\propto\ket{+_x}U_0^\dagger \mathcal{D}\left(G_n+G_n^\dagger\right)\mathcal{D}^\dagger\ket{0}$. Choosing $n=2$ (i.e. squeezing) with a targeted amplitude $\gamma_2=t_f g_2=1/2$, with $g=\omega/10$ and $\epsilon_{0,1}=\omega/4$, we find $\langle a^\dagger a\rangle\approx 2/3$ with a mana $M(\rho)\approx 1/2$. For $n=3$ and $\gamma_3=1/10$, we obtain a cat-like superposition of tri-squeezed states with energy $\langle a^\dagger a\rangle\approx1/10$ and mana $M(\rho)\approx 3/20$. Note that this cat-like superposition contains similar negativity, i.e. mana, than the state $\ket{\gamma_3}$ with $\gamma_3=1/10$ but at lower energy ($\langle a^\dagger a\rangle\approx 2/10$ for the tri-squeezed state $\ket{\gamma_3}$).

\section{Sequence to generate a cubic-phase gate}
As commented in the main text, defining $H_{1s}=ig_1(a^\dagger-a)$, $H_{2s}=ig_2((a^\dagger)^2-a^2)$ and $H_4=g_4(a^4+(a^\dagger)^4)$, we find
\begin{align}
    [H_{1s},[H_4,H_{2s}]]=8 g_1 g_2 g_4(a+a^\dagger)^3.
\end{align}
That is, we exploit the non-Gaussian interaction $H_4$ to obtain other non-Gaussian gates, in this case a term to implement a cubic-phase gate. From the relation
\begin{align}
    e^{iAt}e^{iBt}e^{-iAt}e^{-iBt}=e^{-[A,B]t^2}+O(t^3),
\end{align}
and defining $U_{1s}=e^{-i\tau_1 H_{1s}}$, $U_{2s}=e^{-i\tau_2 H_{2s}}$ and $U_{4}=e^{-i\tau_4 H_{4}}$, it follows
\begin{align}
    U_{4}^\dagger U_2^\dagger U_4 U_2 \approx e^{-[H_4,H_{2s}]\tau_4 \tau_2}=e^{i \tilde{t} C} 
\end{align}
for $\tau_1 g_1\tau_2 g_2 \ll 1$, where $C=i[H_4,H_{2s}]$ and $\tilde{t}=\tau_2\tau_4$. Hence,
\begin{align}
    U_1^\dagger e^{i \tilde{t} C} U_1 e^{-i\tilde{t} C}\approx e^{-[H_{1s},C]\tau_1\tilde{t}}=e^{-i\tau_1\tau_2\tau_4[H_{1s},[H_4,H_{2s}]]}=e^{-i\delta (a+a^\dagger)^3}
\end{align}
with $\delta=8\tau_1 \tau_2 \tau_4 g_1 g_2 g_4$, and $\delta \ll 1$.  The block sequence to generate a cubic-phase gate reads as
\begin{align}\label{eqSM:Ublock}
    U_{\rm block}=U_1 U_{4}^\dagger U_2^\dagger U_4 U_2 U_1^\dagger U_2^\dagger U_{4}^\dagger U_2 U_4\approx e^{-i\delta (a+a^\dagger)^3}.
\end{align}
Concatenating $N$ times this sequence, $(U_{\rm block})^N$, one effectively implements a cubic-phase gate with a {\em cubicity} $N\delta$. 

All the unitaries required to implement a $U_{\rm block}$ in Eq.~\eqref{eqSM:Ublock} can be obtained from Eq.~(5) of the main text, $H_{\rm n-phot}=g_n \sigma_x(a^ne^{-i\phi}+(a^\dagger)^n e^{i\phi})$. In particular, $U_1$, $U_2$ and $U_4$ follow from $n=1$, $n=2$ and $n=4$ with phases $\phi=\pi/2$, $\pi/2$ and $0$, respectively, and provided the two-level system is prepared in $\ket{+_x}$ state. The conjugate-transposed unitaries $U_1^\dagger$, $U_2^\dagger$ and $U_4^\dagger$, can be implemented either by rotating the two-level state $\ket{+_x}\rightarrow\ket{-_x}$ (upon a $\pi$-pulse around the $z$-axis) or by changing the phases of the driving fields $\phi\rightarrow\phi+\pi$.

\end{document}